\documentclass[twocolumn,prl,preprintnumbers,tightenlines,footinbib,superscriptaddress]{revtex4}

\pdfoutput=1
\usepackage{amsmath,amssymb,amsfonts}
\usepackage{graphicx}
\usepackage{booktabs}
\usepackage{hyperref}
\usepackage{color}
\usepackage[sort&compress]{natbib}
\usepackage{tikz}
\usepackage{cleveref}
\newcounter{qnumber}

\newcommand{\drawsquare}[2]{\hbox{%
\rule{#2pt}{#1pt}\hskip-#2pt
\rule{#1pt}{#2pt}\hskip-#1pt
\rule[#1pt]{#1pt}{#2pt}}\rule[#1pt]{#2pt}{#2pt}\hskip-#2pt
\rule{#2pt}{#1pt}}

\newcommand{\Yfund}{\raisebox{-.5pt}{\drawsquare{6.5}{0.4}}}
\newcommand{\Ysymm}{\raisebox{-.5pt}{\drawsquare{6.5}{0.4}}\hskip-0.4pt%
        \raisebox{-.5pt}{\drawsquare{6.5}{0.4}}}
\newcommand{\Ythrees}{\raisebox{-.5pt}{\drawsquare{6.5}{0.4}}\hskip-0.4pt%
          \raisebox{-.5pt}{\drawsquare{6.5}{0.4}}\hskip-0.4pt%
          \raisebox{-.5pt}{\drawsquare{6.5}{0.4}}}
\newcommand{\Yfours}{\raisebox{-.5pt}{\drawsquare{6.5}{0.4}}\hskip-0.4pt%
          \raisebox{-.5pt}{\drawsquare{6.5}{0.4}}\hskip-0.4pt%
          \raisebox{-.5pt}{\drawsquare{6.5}{0.4}}\hskip-0.4pt%
          \raisebox{-.5pt}{\drawsquare{6.5}{0.4}}}
\newcommand{\Yasymm}{\raisebox{-3.5pt}{\drawsquare{6.5}{0.4}}\hskip-6.9pt%
        \raisebox{3pt}{\drawsquare{6.5}{0.4}}}

\newcommand{\Yadjoint}{\raisebox{-3.5pt}{\drawsquare{6.5}{0.4}}\hskip-6.9pt%
        \raisebox{3pt}{\drawsquare{6.5}{0.4}}\hskip-0.4pt
        \raisebox{3pt}{\drawsquare{6.5}{0.4}}}
\newcommand{\Ysquare}{\raisebox{-3.5pt}{\drawsquare{6.5}{0.4}}\hskip-0.4pt%
        \raisebox{-3.5pt}{\drawsquare{6.5}{0.4}}\hskip-13.4pt%
        \raisebox{3pt}{\drawsquare{6.5}{0.4}}\hskip-0.4pt%
        \raisebox{3pt}{\drawsquare{6.5}{0.4}}}
\newcommand{\Ytwoone}{\raisebox{-3.5pt}{\drawsquare{6.5}{0.4}}\hskip-6.9pt%
        \raisebox{3pt}{\drawsquare{6.5}{0.4}}\hskip-0.4pt%
        \raisebox{3pt}{\drawsquare{6.5}{0.4}}\hskip-0.4pt%
        \raisebox{3pt}{\drawsquare{6.5}{0.4}}}%
\newcommand{\Yother}{\raisebox{3pt}{\raisebox{-25.1pt}{\drawsquare{6.5}{0.4}}\hskip-4.7pt
	\raisebox{-15.3pt}{\vdots}\hskip-4.7pt
	\raisebox{-3.5pt}{\drawsquare{6.5}{0.4}}\hskip-6.9pt
    \raisebox{3pt}{\drawsquare{6.5}{0.4}}\hskip-0.4pt
    \raisebox{3pt}{\drawsquare{6.5}{0.4}}\hskip-0.4pt
    \raisebox{3pt}{\drawsquare{6.5}{0.4}}}}
\begin{document}

\title{More Exact Results on Chiral Gauge Theories: the Case of the Symmetric Tensor}

\author{Csaba Cs\'aki}
\email{ccsaki@gmail.com}
\affiliation{Department of Physics, LEPP, Cornell University, Ithaca, NY 14853, USA}

\author{Hitoshi Murayama}
\email{hitoshi@berkeley.edu, hitoshi.murayama@ipmu.jp, Hamamatsu Professor}
\affiliation{Department of Physics, University of California, Berkeley, CA 94720, USA}
\affiliation{Kavli Institute for the Physics and Mathematics of the
  Universe (WPI), University of Tokyo,
  Kashiwa 277-8583, Japan}
\affiliation{Ernest Orlando Lawrence Berkeley National Laboratory, Berkeley, CA 94720, USA}

\author{Ofri Telem}
\email{t10ofrit@gmail.com}
\affiliation{Department of Physics, University of California, Berkeley, CA 94720, USA}
\affiliation{Ernest Orlando Lawrence Berkeley National Laboratory, Berkeley, CA 94720, USA}

\begin{abstract} 
We study dynanics of $SU(N-4)$ gauge theories with fermions in rank-2 symmetric tensor and $N$ anti-fundamental representations, by perturbing supersymmetric theories with anomaly-mediated supersymmetry breaking. We find the $SU(N)\times U(1)$ global symmetry is dynamically broken to $SO(N)$ for $N\geq 17$, a different result from conjectures in the literature. For $N<17$, theories flow to infrared fixed points. 
\end{abstract}

\maketitle

\section{Introduction}
 Understanding the dynamics of strongly coupled chiral gauge theories remains a difficult challenge, as we are lacking effective tools to study them. Even progress on lattice simulations has been slow, since they are hampered by the notorious doubling problem. So far the best we can do is to resort to educated guesses, based on some general guiding principles. The most well-known such framework is ``tumbling" \cite{Raby:1979my,Dimopoulos:1980hn}, where one tries to find the most attractive channel (MAC) among the various fermions, and postulate condensates that will successively break the gauge symmetry till one arrives at a non-chiral QCD-like theory. Tumbling has indeed been used to propose a plausible vacuum structure for the simplest chiral gauge theories: $SU(N)$ with a Weyl fermion in the antisymmetric (or in the symmetric) representation of the gauge group, as well as anti-fundamentals to cancel the gauge anomalies. For both examples the proposal contains a symmetry breaking pattern
that satisfies non-trivial 't Hooft anomaly matching conditions, hence appears to be passing some very non-trivial checks. Recently, these proposals have undergone further scrutiny  in \cite{Bolognesi:2020mpe,Bolognesi:2021yni}, by applying new discrete anomaly matching conditions \cite{Csaki:1997aw} involving the center symmetry $Z_n$ of the gauge group, in the spirit of \cite{Gaiotto:2017yup,Tanizaki:2018wtg,Bolognesi:2019fej}. We do not elaborate more on these generalized consistency conditions, since they seem to automatically hold for our proposal of the IR dynamics, due to its continuous connection to the supersymmetric theory.

The study of the supersymmetric (SUSY) versions of these theories opens up a new method for finding candidate vacua for chiral gauge theories. SUSY - thanks in most part due to holomorphy - allows  for a much greater control of the IR dynamics, and together with anomaly matching often  enables us  to fully pin down the vacuum structure of the theory. The obvious challenge then is to be able to deduce results for the non-SUSY theory by perturbing the SUSY results in a controlled manner. While several attempts along these lines were initiated in the 90's~\cite{Aharony:1995zh,Cheng:1998xg,ArkaniHamed:1998wc}, recently a UV-insensitive method for perturbing SUSY dualities based on anomaly mediation~\cite{Randall:1998uk,Giudice:1998xp} has been proposed in~\cite{Murayama:2021xfj}. Applying the AMSB method to chiral gauge theories (wherever exact results for the supersymmetric limit is known) will always produce a candidate vacuum structure that will automatically satisfy all consistency conditions, since it will be continuously connected to the vacuum of the SUSY theory without the AMSB perturbations, yielding a well-defined procedure for generating a candidate vacuum solution to the non-SUSY chiral gauge theories. Of course this does not automatically imply that we have found the correct ground state of the theory, since one can not rule out the possibility of a phase transition as the SUSY breaking mass terms are raised above the strong coupling scale. Nevertheless we find it significant that at least a plausible conjecture can be formulated this way, which eventually will be compared to the results from the lattice simulations. It may even be the case that holomorphy prohibits a phase transition \cite{Csaki:2021xhi}.

Recently we have used this method to examine the simplest chiral gauge theories based on $SU(N)$ with a rank-2 antisymmetric representation for $N\geq5$ and have identified the vacuum structure of the resulting non-SUSY chiral theory~\cite{Csaki:2021xhi}. We have found that the global symmetry breaking pattern is in fact different than initially conjectured based on tumbling: for odd $N$ we found that the global $SU(N-4)\times U(1)$ symmetry is broken to $Sp(N-5)\times U(1)$, while for the even case to $Sp(N-4)$. While this symmetry breaking pattern did not agree with the original predictions from tumbling, we have explained that assuming additional condensates in the second most attractive channel will fully resolve the discrepancy. 

In this paper we extend our previous work to examine the other examples of non-SUSY chiral gauge theories for which a simple prediction for the vacuum structure based on tumbling exists: the case of $SU(N-4)$ with a fermion $S$ in the rank-2 symmetric representation and $N$ anti-fundamentals $\bar{F}$ to cancel the gauge anomalies. There exist two proposals for the symmetry breaking pattern of this model. The tumbling prediction for this case would be a  MAC leading to a condensate of the symmetric and $N-4$ anti-fundamentals, leading to color-flavor locking with an unbroken $SU(N-4)\times SU(4)\times U(1)$ global symmetry. The anomalies are matched by a composite fermion $S\bar{F}_i \bar{F}_j$ antisymmetric in the $i,j$ flavor indices. Another interesting option is that the entire group confines without breaking any of the global symmetries via condensates, since the same fermion composite actually matches the 't Hooft anomalies of the entire $SU(N)\times U(1)$ global symmetry. 

Similar to the case of the anti-symmetric tensor, we will show that the AMSB method results in a prediction different from either of these two scenarios. The details of the analysis for the symmetric case turn out to be quite different from that of the antisymmetric, since here we have to make use of the Seiberg dual found by Pouliot and Strassler~\cite{Pouliot:1995sk} in terms of a magnetic $Spin(8)$ group.  We find that for $N\geq  17$ the remaining global symmetry is only $SO(N-4)$, and no massless composite fermion is needed in the absence of 't Hooft anomalies. On the other hand for $N<17$ the theory flows to a conformal fixed point in the IR. At the fixed point the supersymmetry breaking terms all vanish, and one is left with a genuine superconformal theory.  

The paper is organized as follows. We first briefly review the anomaly mediation of supersymmetry breaking (AMSB), and the SUSY limit of the $SU(N-4)$ gauge theories with a symmetric tensor and $N$ anti-fundamentals. Then we combine them to find consistent vacua of the non-SUSY theories that can be extrapolated to decouple supersymmetry. We show that we can understand the symmetry breaking pattern \`a la tumbling, even though we need to rely on fermion bilinear condensates that are not in the MAC. The case of infrared fixed points are discussed in the end.

\section{Anomaly Mediation}

In scenarios with anomaly-mediated supersymmetry breaking (AMSB), supersymmetry is broken in a sequestered sector, and is mediated to the visible sector via the superconformal anomaly. The magnitude of the breaking is given by a single number $m$, which enters both at tree and at loop level. The tree-level contribution to the scalar potential is derived from the superpotential,
\begin{align}
	{\cal L}_{\rm tree} &= m \left( \phi_{i} \frac{\partial W}{\partial \phi_{i}} - 3 W \right)
	+ c.c. \label{eq:AMSBW}
\end{align}
In addition, there is loop-level supersymmetry breaking, which generates in tri-linear couplings, scalar masses, and gaugino masses \cite{Pomarol:1999ie},
\begin{align}
	A_{ijk} (\mu) &= - \frac{1}{2} (\gamma_{i} + \gamma_{j} + \gamma_{k})(\mu) m, \label{eq:A} \\
	m_{i}^{2}(\mu) &= - \frac{1}{4} \dot{\gamma}_{i}(\mu) m^{2}, \label{eq:m2} \\
	m_{\lambda}(\mu) &= - \frac{\beta(g^{2})}{2g^{2}}(\mu) m. \label{eq:mlambda} 
\end{align}
Here, $\gamma_{i} = \mu\frac{d}{d\mu} \ln Z_{i}(\mu)$, $\dot{\gamma} = \mu \frac{d}{d\mu} \gamma_{i}$, and $\beta(g^{2}) = \mu \frac{d}{d\mu} g^{2}$.

\section{$SU(N-4)$ with $S$ and $(N)\,\bar{F}_i$  }
First we present a summary of the duality explored in \cite{Pouliot:1995sk}, which will be the basis of our explorations. We consider a supersymmetric $SU(N-4)$ gauge theory with a rank-2 symmetric tensor $S$ and antifundamentals $\bar{F}_i$ ($i = 1, \cdots, N$). For $N\leq 16$, it has an interacting IR fixed point, while for $N\geq 17$, it is in a free magnetic phase. The theory has a global $SU(N)\times U(1)\times U(1)_R$ symmetry, under which the charges of the matter fields are shown in Table~\ref{tab:PCelectric}.

The $SU(N-4)$ theory (henceforth the `electric' theory) has a magnetic dual, which is a non-chiral $Spin(8)$ gauge theory (the double cover of $SO(8)$) with $N$ vectors $q^i$, a spinor $p$ and the $Spin(8)$ singlets $M_{ij} = S \bar{F}_i \bar{F}_j$ and $U = {\rm det}S$. The magnetic theory has a tree-level superpotential:
\begin{equation}
\tilde{W}_{\text{tree}}~=~\frac{1}{\mu^{2}_1}\,M_{ij} q^i q^j\,+\,\frac{1}{\mu^{N-5}_2}\,Upp\,, 
\end{equation}
where the scales $\mu_{1,2}$ are related to the electric (magnetic) strong scales $\Lambda\,(\tilde{\Lambda})$ by 
\begin{equation}
(\Lambda^{2N-11})^2\tilde{\Lambda}^{17-N}~=~\mu^{2N}_1\mu^{N-5}_2\,. 
\end{equation}
The 't Hooft anomaly matching conditions are satisfied and imply that the fields $M$ and $U$ have regular K\"ahler potential at the origin. For later convenience, we switch to canonically normalized fields $\tilde{M}$ and $\tilde{U}$ and introduce the Yukawa couplings $y_{M,U}$,
\begin{equation}\label{eq:treesp}
\tilde{W}_{\text{tree}}~=~y_M\,\tilde{M}qq\,+\,y_U\,\tilde{U}pp\,. 
\end{equation}

\begin{table}[t]
	\centerline{
	\begin{tabular}{|c|c|c|c|c|} \hline
	& $SU(N-4)$ & $SU(N)$&$U(1)$&$U(1)_R$ \\ \hline
	$S$ & $\Ysymm$ &$\bf1$ &$-2N$&$\frac{12}{(N+1)(N-4)}$\\ \hline
	$\bar{F}_i$ & $\overline{\Yfund}$ & $\Yfund$&$2N-4$&$\frac{6 (N-5)}{(N+1)(N-4)}$ \\ \hline \hline
	 $M_{ij}=S\bar{F}_i\bar{F}_j$&$\bf1$  &$\Ysymm$&$2N-8$&$\frac{12}{N+1}$\\ \hline
	 $U=\text{det}S$&$\bf1$  &$\bf 1$&$2N(4-N)$&$\frac{12}{N+1}$\\ \hline
	\end{tabular}
	}
	\caption{Particle content of the electric $SU(N-4)$  theory. We omit the baryons $B,\,B_n$ since they are not dynamical and don't play a role in our analysis.}\label{tab:PCelectric}
%
\vspace{10pt}
	\centerline{
	\begin{tabular}{|c|c|c|c|c||c|} \hline
	& $Spin(8)$ & $SU(N)$&$U(1)$&$U(1)_R$&$SO(N)$ \\ \hline
	$q^i$ &$\bf8_v$ &$\bar{\Yfund}$ &$4-N$&$\frac{N-5}{N+1}$&$\Yfund$\\ \hline
	$p$ & $\bf8_s$ &$\bf1$ &$N(N-4)$&$\frac{N-5}{N+1}$&$\bf1$ \\ \hline
	$M_{ij}$ & $\bf1$ &$\Ysymm$ &$2N-8$&$\frac{12}{N+1}$&$\bf1+\Ysymm$ \\ \hline
	$U$ & $\bf1$ &$\bf 1$ &$2N(4-N)$&$\frac{12}{N+1}$&$\bf1$ \\ \hline
	\end{tabular}
	}
	\caption{Particle content of the magnetic $Spin(8)$  theory. We omit the baryons $b,\,b_n$ since they are not dynamical don't play a role in our analysis. Note that we use the same name for $M_{ij},\,U$ as in the electric theory, due to their indentical representations.}\label{tab:PCmagnetic}
\end{table}
The duality between the $SU(N-4)$ and $Spin(8)$ also maps the composite operators of the theory: $S\bar{F}_i\bar{F}_j\leftrightarrow M_{ij}$ and $\text{det}\,S\leftrightarrow U$. There are also gauge invariant baryonic operators on both the electric and the magnetic side, but they do not play a role in the dynamics below and so we do not discuss them further in this paper.

\section{Adding AMSB}

Perturbing the duality via the AMSB mechanism will result on the electric side in positive scalar masses as well as gaugino masses, leaving in the IR the non-supersymmetric chiral gauge theory of interest. We will then have to identify the effect of the AMSB on the magnetic $Spin(8)$ theory and find the global minimum of its supersymmetry breaking potential. We first focus on the case $N\geq 17$, in which the theory is in the free magnetic phase. A naive local minimum is obtained by directly adding the AMSB to the tree level potential \eqref{eq:treesp}. In this case the tree-level AMSB contribution from \eqref{eq:AMSBW} vanishes, and so the supersymmetry breaking is generated by the loop level $A$-terms \eqref{eq:A} and soft masses (\ref{eq:m2},\ref{eq:mlambda}), leading to a local minimum at
\begin{align}\label{eq:notglobal}
V\,&\approx\,-\left(\frac{\lambda^2}{16\pi^2}\right)^4\,m^4\,,
\end{align}
where $\lambda$ is some $\mathcal{O}(1)$ combination the gauge and superpotential couplings $g,\,y_M,$ and $y_U$. While this is indeed a local minimum of the potential, we will now show, this is not the global minimum of the theory.

To find the global minimum, we first consider the magnetic theory on the moduli space by turning on $\left\langle M \right\rangle$ of rank $N$, as well as $\left\langle U \right\rangle$. As we shall see, the global minimum end up being at small values of these VEVs, justifying our weakly coupled analysis in the (initially) IR free magnetic theory. Once $q_i,p$ are integrated out, the magnetic theory becomes asymptotically free again: indeed the IR theory is a pure $Spin(8)$ SYM, with a scale 
\begin{eqnarray}
\tilde{\Lambda}_{L}^{18}\,=\,\frac{\text{det}\tilde{M}\,\tilde{U}}{\tilde{\Lambda}^{N-17}}\,.
\end{eqnarray}
It develops a gaugino condensate with a dynamically generated superpotential \cite{Intriligator:1995id}\footnote{Here and below, we absorb scheme-dependent numerical constants \cite{Finnell:1995dr} into the definition of the scale $\tilde{\Lambda}_L$ which does not affect any of the discussions below.}
\begin{equation}\label{eq:Wdyn}
W_{\text{dyn}}\,=\,e^{i\frac{\pi k}{3}}(\tilde{\Lambda}^{18}_{L})^{1/6}\,
=\,e^{i\frac{\pi k}{3}}\left(\frac{\text{det}\tilde{M}\,\tilde{U}}{\tilde{\Lambda}^{N-17}}\right)^{1/6}\,,
\end{equation}
where $k=0,\ldots,5$ denoting six different vacua, originating from taking the sixth root in the gaugino condensate. Taking this dynamical superpotential into account, and using \eqref{eq:AMSBW}, we find the tree-level AMSB contribution
\begin{align}
	{\cal L}_{\rm tree} &= m \frac{N-17}{6}\,W_{\text{dyn}}
	+ c.c. \label{eq:AMSBWmag}
\end{align}
The loop-level AMSB terms (\ref{eq:A},\ref{eq:m2},\ref{eq:mlambda}) are negligible with respect to this tree-level contribution. The scalar potential from the superpotential \eqref{eq:Wdyn} and the tree-level AMSB contribution \eqref{eq:AMSBWmag} has a supersymmetry breaking minimum at
\begin{align}\label{eq:min}
	\tilde{M}_{ij}\,&\approx\, \delta_{ij} m \left( \frac{\tilde{\Lambda}}{m} \right)^{\frac{N-17}{N-11}},
	\quad \tilde{U}\,\approx\,m \left( \frac{\tilde{\Lambda}}{m} \right)^{\frac{N-17}{N-11}}, \nonumber\\
    V\,&\approx\,-m^4\,{\left(\frac{\tilde{\Lambda}}{m}\right)}^{\frac{2(N-17)}{N-11}}\,.
\end{align}
Indeed, for $N>17$ this minimum is deeper than the one in \eqref{eq:notglobal}. We will come back to the marginal case $N=17$ shortly. 
On the complex plane of $z=({\rm det}\tilde{M}\ \tilde{U})$, six branches are connected through branch cuts from one Riemann sheet to another. The minimum of the potential finds itself on the sheet where $m W_{\rm dyn}$ in Eq.~\eqref{eq:AMSBWmag} can be real and negative.
We recall that the magnetic theory is IR free for $N\geq 17$ and in this case the VEVs are both much smaller than the Landau pole and much larger than the scale $\Lambda_L$ of the gaugino condensate.  For this reason the minimum eq.~\eqref{eq:min} is at weak coupling. 

At the minimum the global $SU(N)\times U(1)$ symmetry is broken to $SO(N)$, (the $U(1)_R$ is explicitly broken by the AMSB) and all fermions become massive. There are $\frac{1}{2}N(N+1)$ Nambu-Goldstone bosons corresponding to the breaking, and none of them are eaten since the $Spin(8)$ gauge symmetry remains unbroken. Since $SO(N)$ is anomaly free, there are no non-trivial 't Hooft anomaly matching conditions. In the IR theory, the $M_{ij}$ and $U$ all play the role of Goldstone superfields, each one containing exactly one of the $\frac{1}{2}N(N+1)$ Nambu-Goldstone boson for the $SU(N)\times U(1)/SO(N)$ breaking, except for one combination of $U$ and ${\rm Tr}M$ which is the direction of the potential and is not a NGB.

Note that the VEV $M_{ij}$ has the full rank $N$ in the magnetic theory, while in the electric theory their maximal rank is $N-4$. This is completely consistent with the duality: in the supersymmetric limit, the rank condition is enforced on $M_{ij}$ \textit{dynamically} \cite{Intriligator:1995au}, because the superpotential Eq.~\eqref{eq:Wdyn} requires $\tilde{M}^{-1}{\rm det}\tilde{M}\ \tilde{U}={\rm det}\tilde{M}=0$. Therefore, along $F$-flat directions corresponding to potential SUSY preserving minima ${\rm rank}\, \tilde{M}_{ij}<N$. Directions with maximal rank satisfying ${\rm det}\tilde{M}\ \tilde{U} \neq 0$ (as we assumed in deriving Eq.~\eqref{eq:Wdyn}) necessarily correspond to SUSY breaking vacua. In our case  AMSB indeed stabilizes the minimum \eqref{eq:min} along the direction where $\langle \tilde{M}_{ij}\rangle$ has maximal rank, and away from the classically expected rank condition.  This situation is similar to the way the  full-rank meson field is found in the ISS model when $N_c < N_f$ in the electric theory \cite{Intriligator:2006dd}, with the mass perturbation  stabilizing the minimum away from the classical rank condition.

The marginal case $N=17$ is a little subtle. In this case the one-loop beta function of the magnetic gauge coupling vanishes, though the two-loop beta function does not and is IR free. Consequently, the theory is still in the free magnetic phase. While in this case the minima \eqref{eq:min} and \eqref{eq:notglobal} have the same dependence on $m$, the minimum \eqref{eq:notglobal} is much shallower, since it depends on the slow 2-loop running of the gauge coupling, and quasi-IR-fixed point behavior of the Yukawa couplings that tracks the gauge coupling. Then \eqref{eq:min} remains the global minimum for $N=17$. We also verified that the two-loop mass-squared for $\tilde{M}$ and $\tilde{U}$ is smaller than the square of the one-loop $A$-term, and the origin is unstable. Therefore we find $\tilde{M} \sim \tilde{U}\sim \frac{\lambda^2}{16\pi^2} m \neq 0$, and the symmetry breaking pattern is the same as the rest of $N\geq 17$ cases. 

 So far the results obtained are exact. As in previous analyses with the AMSB \cite{Murayama:2021xfj,Csaki:2021xhi}, we now take the limit $m\rightarrow\infty$ to extrapolate to the non-supersymmetric theory. Though there might be a phase transition on the way, our analysis yields a plausible conjecture for the IR behavior of non-supersymmetric $SU(N-4)$ with a symmetric $S$ and $N$ anti-fundamentals. As a continuous limit of a self consistent supersymmetric analysis, our method is guaranteed to fulfill all 't Hooft matching conditions, including generalized ones \cite{Gaiotto:2017yup,Tanizaki:2018wtg,Bolognesi:2019fej,Bolognesi:2020mpe,Bolognesi:2021yni}. Note also that in a complete AMSB model our SUSY breaking spurion $m$ (the $F$-component of the compensator) would originate from the constant term in the superpotential leading to a relation $ m  = \frac{1}{M_{\rm Pl}^2}W_0$, where $W_0$ is the superpotential in the SUSY breaking sector.  If there is a phase transition while increasing the amount of SUSY breaking it would happen for $|m|\approx |\Lambda|$, corresponding to the condition $|\Lambda M_{\rm Pl}^2|^2 \approx |W_0|^2$. Since both $\Lambda$ and $W_0$  are chiral superfields (or products thereof), any relation among them defining the phase boundaries should be given by a holomorphic expression. Consequently, the phase boundary should be even dimensional: either isolated points or the entire complex plane. Since the latter is implausible it would have to be isolated points. However those can not corresponds to the relation $|\Lambda M_{\rm Pl}^2|^2 \approx |W_0|^2$ which would imply the phase boundary to be a circle on the complex $W_0$ plane.  This suggests that such a phase boundary should not exist. It would be interesting to see if this argument can be made more rigorous.

\begin{table}[t]
	\centerline{
	\begin{tabular}{|c|c|c|c|c|c|} \hline
	\raisebox{-5pt}{\,$\text{Irrep}\,$}&
	\raisebox{-5pt}{$\Yfund$}&
	\raisebox{-5pt}{$\Yasymm$}&
	\raisebox{-5pt}{$\Ysymm$}&
    \raisebox{-5pt}{$\Ysquare$}&
    \raisebox{-5pt}{$\Ytwoone$}
\\[10pt]\hline
	\raisebox{-5pt}{$C_2$}&
	\raisebox{-5pt}{$\frac{n^2-1}{2n}$}&
	\raisebox{-5pt}{$\frac{(n+1)(n-2)}{n}$}&
	\raisebox{-5pt}{$\frac{(n-1)(n+2)}{n}$}&
	\raisebox{-5pt}{$\frac{2(n^2-4)}{n}$}&
	\raisebox{-5pt}{$\frac{2(n^2+n-4)}{n}$}\\[10pt]\hline
	\end{tabular}}
\vspace{10pt}
	\begin{tabular}{|c|c|c|} \hline
	\raisebox{-16pt}{$\,\text{Irrep}\,$} &
	\raisebox{-16pt}{$\Yfours$}&
	\raisebox{-7pt}{$\,\,\Yother\,$}\\[30pt]
\hline
\raisebox{-5pt}{$C_2$}&
\raisebox{-5pt}{$\,\,\frac{2(n+4)(n-1)}{n}\,\,$}&
\raisebox{-5pt}{$\,\,\frac{n^2-6n^2-7n+80}{2n}\,\,$}
\\[10pt]\hline
	\end{tabular}
	\caption{Quadratic Casimirs for $SU(n)$. The Casimirs are the same for irreps and their conjugates. For the last one, there are $n-1$ boxes vertically for the first column.}\label{tab:SU(n)Casimirs}
%
\vspace{10pt}
	\centerline{
	\begin{tabular}{|c|c|c|} \hline
	\raisebox{-1pt}{\,Constituents\,} &
	\raisebox{-1pt}{\,Channel\,}&
	\raisebox{-1pt}{$\,C_2(\text{channel})-C_2(1)-C_2(2)\,$}\\[5pt] \hline
	\raisebox{-2pt}{$S,\,\bar{F}$} &
	\raisebox{-2pt}{$\Yfund$}&
	\raisebox{-2pt}{$-\frac{(n-1)(n+2)}{n}$} \\[6pt]\hline
	\raisebox{-3pt}{$S,\,S$} &
	\raisebox{-3pt}{$\Ysquare$}&
	\raisebox{-3pt}{$-\frac{2(n+2)}{n}$}\\[6pt] \hline
    \raisebox{-3pt}{$\bar{F},\,\bar{F}$} &
    \raisebox{-3pt}{$\Yasymm$}&
    \raisebox{-3pt}{$-\frac{n+1}{n}$}\\[6pt] \hline
    \raisebox{-3pt}{$S,\,S$} &
    \raisebox{-3pt}{$\Ytwoone$}&
    \raisebox{-3pt}{$-\frac{4}{n}$} \\[6pt] \hline
    \raisebox{-3pt}{$\bar{F},\,\bar{F}$} &
    \raisebox{-3pt}{$\Ysymm$}&
    \raisebox{-3pt}{$\frac{n-1}{n}$} \\[6pt] \hline
	\raisebox{-3pt}{$S,\,S$} &
	\raisebox{-3pt}{$\Yfours$}&
	\raisebox{-3pt}{$\frac{4(n-1)}{n}$} \\[6pt] \hline
	\raisebox{-14pt}{$S,\,\bar{F}$} &
	\raisebox{-7pt}{$\,\,\Yother\,$}&
	\raisebox{-14pt}{$\frac{n^2-9n^2-9n+85}{2n}$} \\[30pt]\hline
	\end{tabular}
	}
	\caption{$SU(n)$ Channels, ordered by most to least attractive. Note that in our case $n=N-4$.}\label{tab:SU(n)Channels}
\end{table}

\begin{table}[t]
	\centerline{
	\begin{tabular}{|c|c|c|c|c|c|} \hline
	\raisebox{-5pt}{\,$\text{Irrep}\,$}&
	\raisebox{-5pt}{$\,\bf 1\,$}&
	\raisebox{-5pt}{$\Yfund$}&
	\raisebox{-5pt}{$\Yasymm$}&
	\raisebox{-5pt}{$\,\Ysymm\,$}&
    \raisebox{-5pt}{$\Ysquare$}\\[10pt]\hline
	\raisebox{-5pt}{$C_2$}&
	\raisebox{-5pt}{$0$}&
	\raisebox{-5pt}{$\,\frac{(n-1)}{2}\,$}&
	\raisebox{-5pt}{$n-2$}&
	\raisebox{-5pt}{$n$}&
	\raisebox{-5pt}{$\,2n-2\,$}\\[10pt]\hline
	\end{tabular}}
\vspace{10pt}
	\begin{tabular}{|c|c|c|c|c|} \hline
	\raisebox{-5pt}{\,$\text{Irrep}\,$}&
	\raisebox{-5pt}{$\Ythrees$}&
	\raisebox{-5pt}{$\,\Yfours\,$}&
	\raisebox{-5pt}{$\Yadjoint$}&
    \raisebox{-5pt}{$\Ytwoone$}
\\[10pt]\hline
	\raisebox{-5pt}{$C_2$}&
	\raisebox{-5pt}{$\,\frac{3(n+1)}{2}\,$}&
	\raisebox{-5pt}{$2n+4$}&
	\raisebox{-5pt}{$\frac{3(n-1)}{2}$}&
	\raisebox{-5pt}{$2n$}\\[10pt]\hline
	\end{tabular}
	\caption{Quadratic Casimirs for $SO(n)$.}\label{tab:SO(n)Casimirs}
%
\vspace{10pt}
	\centerline{
	\begin{tabular}{|c|c|c|} \hline
	\raisebox{-1pt}{\,Constituents\,} &
	\raisebox{-1pt}{\,Channel\,}&
	\raisebox{-1pt}{$\,C_2(\text{channel})-C_2(1)-C_2(2)\,$}\\[5pt] \hline
	\raisebox{-2pt}{$S,\,S$} &
	\raisebox{-2pt}{$\bf 1$}&
	\raisebox{-2pt}{$-2n$} \\[6pt]\hline
	\raisebox{-3pt}{$S,\,S$} &
	\raisebox{-3pt}{$\Yasymm$}&
	\raisebox{-3pt}{$-n-2$}\\[6pt] \hline
    \raisebox{-3pt}{$S,\,\bar{F}$} &
    \raisebox{-3pt}{$\Yfund$}&
    \raisebox{-3pt}{$-n$}\\[6pt] \hline
    \raisebox{-3pt}{$S,\,S$} &
    \raisebox{-3pt}{$\Ysymm$}&
    \raisebox{-3pt}{$-n$} \\[6pt] \hline
    \raisebox{-3pt}{$\bar{F},\,\bar{F}$} &
    \raisebox{-3pt}{$\bf 1$}&
    \raisebox{-3pt}{$1-n$} \\[6pt] \hline
	\raisebox{-3pt}{$S,\,S$} &
	\raisebox{-3pt}{$\Ysquare$}&
	\raisebox{-3pt}{$-2$} \\[6pt] \hline
	\raisebox{-3pt}{$\bar{F},\,\bar{F}$} &
	\raisebox{-3pt}{$\Yasymm$}&
	\raisebox{-3pt}{$-1$} \\[6pt]\hline
	\raisebox{-3pt}{$S,\,\bar{F}$} &
    \raisebox{-3pt}{$\Yadjoint$}&
    \raisebox{-3pt}{$-1$} \\[6pt] \hline
    \raisebox{-3pt}{$S,\,S$} &
    \raisebox{-3pt}{$\Ytwoone$}&
    \raisebox{-3pt}{$0$} \\[6pt] \hline
    \raisebox{-3pt}{$\bar{F},\,\bar{F}$} &
    \raisebox{-3pt}{$\Ysymm$}&
    \raisebox{-3pt}{$1$} \\[6pt] \hline
    \raisebox{-3pt}{$S,\,\bar{F}$} &
    \raisebox{-3pt}{$\Ythrees$}&
    \raisebox{-3pt}{$2$} \\[6pt] \hline
    \raisebox{-3pt}{$S,\,S$} &
    \raisebox{-3pt}{$\Yfours$}&
    \raisebox{-3pt}{$4$} \\[6pt] \hline
	\end{tabular}
	}
	\caption{$SO(n)$ Channels, ordered by most to least attractive. Note that in our case $n=N-4$.}\label{tab:SO(n)Channels}
\end{table}
\section{Tumbling Interpretation}
Here we would like to interpret our SUSY+AMSB analysis of $SU(N-4)$ with a symmetric $S$ and $N$ $\bar{F}$ in the heuristic tumbling approach \cite{Raby:1979my,Dimopoulos:1980hn}. The unbroken global $SO(N)$ symmetry in the IR hints at a symmetric $\bar{F}_{\{i,}\bar{F}_{j\}}$ condensate which breaks the gauge group $SU(N-4)\rightarrow SO(N-4)$ and the global symmetry $SU(N)\rightarrow SO(N)$. However, a symmetric condensate is not an attractive channel for the $SU(N-4)$ gauge symmetry. The solution to this conundrum is the simultaneous condensation of two channels. The first is
\begin{eqnarray}
\Ysquare: \qquad
S_{ab}S_{cd}-S_{ad}S_{cb}\,\propto\,\delta_{ab}\delta_{cd}-\delta_{ad}\delta_{cb}\,.
\end{eqnarray}
This condensate is attractive in $SU(N-4)$. It breaks the $U(1)$ global symmetry, and higgses the gauge symmetry down to $SO(N-4)$. Under the reduced gauge symmetry, the theory is now vector-like and confines.  The symmetric $\bar{F}_{\{i,}\bar{F}_{j\}}$ is now a color singlet. It is attractive, and condenses as
\begin{eqnarray}
\delta_{ab} \bar{F}^{a}_{i}\bar{F}^{b}_{j}\propto \delta_{ij}
\end{eqnarray}
breaking the global $SU(N)$ symmetry down to $SO(N)$. The candidate Nambu-Goldstone bosons for the $SU(N)/SO(N)$ coset are  $\bar{F}_{\{i,}\bar{F}_{j\}}$, while the Goldstones for the two broken $U(1)$s can be taken to be ${(\bar{F}^a_i)}^2$ and $S^2_{ab}$. 
By examining tables~\ref{tab:SU(n)Channels}-\ref{tab:SO(n)Channels}, we see that the condensates described above are attractive, but they are certainly not the most attractive channel (MAC). Our suggestion for the IR dynamics of the theory is then different from the one suggested by tumbling, or the other proposed phase with fully unbroken global symmetries and no condensates.

\section{Non Abelian Coulomb Phase}
For $N<16$, the theory has an IR fixed point \cite{Pouliot:1995sk}, in which both the electric and magnetic descriptions of the theory are equally useful. In the supersymmetric limit the anomalous dimensions $\gamma_i$ are scale independent and so $\dot{\gamma}_i=0$ and the AMSB soft masses vanish at the IR fixed point. More specifically $\gamma_i=3R_i-2$ where $R_i$ is the $R$-charge. However, in the presence of non-anomalous $U(1)$ symmetries, the definition of the $R$-symmetry is ambiguous. The combination that appears in the super-conformal algebra is fixed by $a$-maximalization \cite{Intriligator:2003jj}, and the resultant $a$ is the Euler trace anomaly coefficient that always decreases with the renormalization-group flow \cite{Komargodski:2011vj} analogous to the Zamolodchikov's $c$-theorem \cite{Zamolodchikov:1986gt}. The Euler trace anomaly coefficient $a$ is defined by \cite{Anselmi:1997am}
\begin{align}
    a &= \frac{3}{32} \left( 3 {\rm Tr} R^3 - {\rm Tr} R \right),
\end{align}
where the trace sums over all fermions in the theory. 
Obviously the definition is common between the electric and magnetic theories as long as the 't Hooft anomaly matching conditions are satisfied. Using the $U(1)$ charge $Q$ in Table~\ref{tab:PCelectric}, we maximize $a$ using the combination $R' = R + t Q$, and find the (local) maximum at
\begin{align}
    t = \frac{9 N^2-24 N+39-(N+1)\sqrt{73 N^2-362 N+433}}{6 (N-4) (N-1) (N+1)}\ .
\end{align}
$t$ is very small, $t\approx 0.0245$ for $N=6$ and even smaller for larger $N$. We can see that $t=0$ for $N=17$, implying that the dimension of $M$ and $U$ become 1 at that point, signaling the beginning of the free magnetic phase, where $M$ and $U$ are indeed free fields.
With the anomalous dimensions $\gamma_i = 3 R'_i - 2$, the anomaly-free condition for $U(1)'_R$ guarantees that the NSVZ beta function vanishes \cite{Anselmi:1997am,deGouvea:1998ft}.


Supersymmetry breaking quickly vanishes as the theory approaches the IR fixed point. For example, we can work out the gaugino mass by expanding the beta function to the first order around the coupling $g_*$ at the fixed point, 
\begin{align}
\beta(g^2)\,&=\,\beta'_{*}\left(g^2-g^2_*\right)\, + O\left(g^2-g^2_*\right)^2,
\end{align}
and so in the vicinity of the IR fixed point ($\beta'_{*}>0$),
\begin{align}
g^2(\mu)\,&=\,g^2_*\,+\,\left[g^2(\mu')-g^2_*\right]{\left(\frac{\mu}{\mu'}\right)}^{\beta'_{*}}\,,
\end{align}
for energy scales $\mu>\mu'>\Lambda_*$ where $\Lambda_*$ is the energy scale of the IR fixed point. Eq.~\eqref{eq:mlambda} then gives the gaugino mass
\begin{align}
m_{\lambda}(\mu) &= -m\, \frac{\beta'_{*}}{2}\frac{g^2(\mu')-g^2_*}{g^2_*+\left(g^2(\mu')-g^2_*\right){\left(\frac{\mu}{\mu'}\right)}^{\beta'_{*}}}{\left(\frac{\mu}{\mu'}\right)}^{\beta'_{*}} \nonumber\\
&\approx m\, \frac{\beta'_{*}}{2}\left[1-\frac{g^2(\mu')}{g^2_*}\right]{\left(\frac{\mu}{\mu'}\right)}^{\beta'_{*}}\,,
\end{align}
where in the last step we assumed $\mu\ll\mu'$ and neglected the power-suppressed second term in the denominator. The gaugino mass as expected is power-law suppressed, and tends quickly to zero as we approach the IR.

\section{Conclusions}

We have identified the IR phase of the non-SUSY chiral $SU(N-4)$ gauge theory with a fermion in the symmetric representation, as well as $N$ anti-fundamentals, obtained via perturbing the relevant SUSY duality with the AMSB. For $N\geq 17$ the theory is confining with the $SU(N)$ global symmetry broken to $SO(N)$ and no massless composite fermions are needed to match anomalies. For $N<17$ the theory flows to a (super-)conformal fixed point, providing another interesting example of a non-SUSY theory flowing to a SUSY fixed point in the IR. 

While the results are obtained in the limit of small SUSY breaking $m\ll \Lambda$, they do provide a plausible candidate vacuum structure for these theories even when $m\gg \Lambda$, satisfying all possible consistency conditions by construction. The resulting vacuum structure obtained with this method differs significantly from either of the two conjectured phases of the non-SUSY theory, and can be given an interesting interpretation in the tumbling framework via two condensates, neither of which would correspond to the MAC (and one of them becomes attractive only in the presence of the first condensate). Whether this is indeed the correct phase of the non-SUSY theory, or if a phase transition occurs at $m\sim \Lambda$ will have to be eventually verified by dedicated lattice simulations. 

\section{Acknowledgments}

\begin{acknowledgments}

We thank Ken Intriligator for careful reading of the manuscript and useful comments. CC is supported in part by the NSF grant PHY-2014071 as well as the US-Israeli BSF grant 2016153. OT and HM were supported in part by the DOE under grant DE-AC02-05CH11231.
HM was also supported in part by the NSF grant
PHY-1915314, by the JSPS Grant-in-Aid for
Scientific Research JP20K03942, MEXT Grant-in-Aid for Transformative Research Areas (A)
JP20H05850, JP20A203, by WPI, MEXT, Japan, and Hamamatsu Photonics, K.K.
\end{acknowledgments}

\bibliographystyle{utcaps_mod}
\bibliography{chiralrefs}

\end{document}